\documentclass[12pt]{article}

\usepackage{cite}
\usepackage{epsf}
\usepackage{graphicx}
\usepackage{psfrag}
\usepackage[dvips]{epsfig}

\usepackage[cp866]{inputenc}
\usepackage[english]{babel}
\usepackage{amssymb,indentfirst}

\makeatletter
\renewcommand \thesection {\@arabic\c@section.}
\renewcommand\thesubsection   {\thesection\@arabic\c@subsection.}
\renewcommand\thesubsubsection{\thesubsection\@arabic\c@subsubsection.}
\makeatother

\def\starup#1{\mbox{$\raise1.8ex\hbox{$*$} \kern-.7em#1$}}
\def\krup#1{\mbox{$\raise1.8ex\hbox{$+$} \kern-1.0em#1$}}
\def\linup#1{\mbox{$\raise1.9ex\hbox{---} \kern-1.0em#1$}}

\begin{document}

\title{
Fermionic decays of scalar leptoquarks and scalar gluons 
in the minimal four color symmetry model
 }

\author{P.Yu.~Popov\thanks{E-mail: popov\_p@univ.uniyar.ac.ru} , \,
 A.V.~Povarov\thanks{E-mail: povarov@univ.uniyar.ac.ru}, \,
 A.D.~Smirnov\thanks{E-mail: asmirnov@univ.uniyar.ac.ru}\\
{\small\it Division of Theoretical Physics, Department of Physics,}\\
{\small\it Yaroslavl State University, Sovietskaya 14,}\\
{\small\it 150000 Yaroslavl, Russia.}}
\date{}
\maketitle

\begin{abstract}
\noindent
Fermionic decays of the scalar leptoquarks 
$ S=S_1^{(+)}, \, S_1^{(-)}, \, S_m $
 and of the scalar gluons  
$F=F_1, \, F_2$
predicted by the four color symmetry model 
with the Higgs mechanism 
of the quark-lepton mass splitting are investigated. 
Widths and branching ratios of these decays 
are calculated and analysed in dependence on coupling constants 
and on masses of the decaying particles. 
It is shown that the decays 
$ S_1^{(+)}\to tl^+_j, \,\,\,S_1^{(-)}\to \nu_i\tilde b, \,\,\,
S_m\to t\tilde \nu_j, \,\,\,  
 F_1\to t\tilde b, \,\,\, F_2\to t\tilde t $
are dominant 
 with the widths of order of a few GeV 
for $m_S,~m_F<1$~TeV and with the total branching ratios  
close to 1. 
In the case of $m_S < m_t$ the  dominant scalar leptoquark decays are 
$ S_1^{(+)}\to~cl_j^+, \, S_1^{(-)}\to~\nu_i\tilde b, \,   
S_m\to~b\l_j^+, \, S_m\to~ c\tilde \nu_j $ 
 with the total branching ratios    
$Br(S_1^{(+)}\to~cl^+) \approx $ $Br(S_1^{(-)}\to~\nu\tilde b)~\approx~1$,     
$Br(S_m\to~bl^+) \approx~0.9$ and $Br(S_m\to~c\tilde \nu)~\approx~0.1.$ 
A search for such decays at the LHC and Tevatron may be of interest. 

\vspace{5mm}


\end{abstract}




One of the goals of the forthcoming experiments at LHC 
(in addition to the search for the Higgs boson and to the further studies 
of the Standard Model (SM)) will be the search for the possible effects 
of the new physics beyond the SM. There is a lot of the variants of new physics 
beyond the SM predicting new effects at energies of LHC 
(supersymmetry, left - right symmetry, two Higgs model, etc.). 

One of such variants can  be the variant induced by the possible 
four color symmetry between quarks and leptons of Pati--Salam type \cite{PS}.
The immediate consequence of this symmetry is the prediction 
of the gauge leptoquarks which, however, occur to be relatively heavy  
(for example, the resulted from the unobservation of the 
$K^0_L \rightarrow \mu^{\pm} e^{\mp}$ 
decays  the most stringent lower mass limit for the vector leptoquarks 
is of order of $10^3$ TeV).
By this reason it is usually thought that the effects of the four color symmetry 
at the colliders energies are too small to be directly detectable 
in the collider experiments of the near future.

It should be noted however that in addition to the gauge leptoquarks 
the four color symmetry can predict also the new scalar particles. 
Thus, in the case of the Higgs mechanism 
of splitting the masses of quarks and leptons the four color symmetry 
in its minimal realization on the gauge group 
\begin{eqnarray}
G=SU_V(4)\times SU_L(2)\times U_R(1) \label{eq:a1}
\end{eqnarray}
(MQLS-model \cite{AD1,AD2}) needs the existence 
of the scalar particles belonging to the 
(15,2,1) - multiplet of the group G. 
This multiplet contains fifteen scalar $SU_{L}(2)$-doublets: two scalar 
leptoquark doublets $S^{(\pm)}$, scalar gluon doublet $F$ and a  
$SU_{c}(3)$-colorless scalar doublet which in admixture 
with the (1, 2, 1) - doublet forms the SM Higgs doublet $\Phi^{(SM)}$ and 
an additional colorless scalar doublet $\Phi'$. 
All these scalar doublets are necessary 
\cite{PovSm1}
for  splitting the the masses of quarks from those of leptons by the Higgs 
mechanism and for generating the quark - lepton mass splittings including 
the so large mass splittings as the $b - \tau $ and $t - \nu_{\tau} $ ones.    

 Because of their Higgs origin the coupling constants of these
scalar doublets with the fermions are proportional to
the ratios $m_{f}/ \eta $ of the fermion masses $m_f$ to the SM
VEV $\eta$. The effects of these scalar leptoquarks in the
processes with the ordinary u-, d-, s- quarks are small because of the
smallness of the corresponding coupling constants 
$ m_u/ \eta \sim m_d/ \eta \sim 10^{-5},  m_s/ \eta \sim 10^{-3}$ 
(but these effects can be significant in c-, b- and, especially,
in top-physics 
$ m_c/ \eta \sim m_b/ \eta \sim 10^{-2} , \, m_t/ \eta \sim 0.7$). 
As a result the scalar leptoquark doublets $S^{(\pm)}$ and 
the scalar gluon doublet $F$  can be relatively light, with
masses below 1 TeV, without any contradictions with the 
$K^0_L \rightarrow \mu^{\pm} e^{\mp}$ data or with the radiative correction 
limits   \cite{AD3,PovSm2}. 
 $SU_{c}(3)$-colored, the scalar leptoquarks $S^{(\pm)}$ and the scalar gluons $F$ 
can be in pairs produced in $pp$-collisions via gluon-gluon fusion and, in part, 
via quark-antiquark annihilation. In the case of the masses below 1 TeV 
the corresponding total cross section 
for the scalar leptoquark pairs \cite{BBK1,BBK2} (and, seemly, for the the scalar gluon ones) 
is known to be sufficient for their effective production at LHC.          
For the detection of these particles it is necessary to know their dominant 
decay modes and the corresponding widths and branching ratios.   

In this paper we calculate and discuss the widths of the fermionic decays 
of the scalar leptoquark and gluon doublets predicted by 
four color symmetry with the Higgs mechanism of splitting the masses of
quarks and leptons in frame of MQLS-model \cite{AD1, AD2}. 
We evaluate and analyze the widths of the dominant decay modes 
in dependence on the masses of the decaying particles.    

In MQLS model the basic left ($L$) and right ($R$) quarks
${Q'}^{L,R}_{ia\alpha}$ and leptons ${l'}^{L,R}_{ia}$
form the fundamental quartets of $SU_{V}(4)$ color group and can
be written, in general, as superpositions
\begin{eqnarray}
{Q'}^{L,R}_{ia\alpha}=\sum_{j}(A^{L,R}_{Q_a})_{ij} \, Q^{L,R}_{ja\alpha} ,
\,\,\,\,\,\,\,
{l'}^{L,R}_{ia}=\sum_{j}(A^{L,R}_{l_a})_{ij} \, l^{L,R}_{ja} \; \label{eq:fmix}
\end{eqnarray}
of the quark and lepton mass eigenstates $Q^{L,R}_{ia\alpha}$ ,
$l^{L,R}_{ia}$, where $i,j=1, \, 2, \, 3$ are the generation indexes,
$ a = 1, 2 $ and $ \alpha = 1, 2, 3 $ are the $ SU_{L}(2) $  and
$ SU_{c}(3) $ indexes,
$Q_{i1} \equiv u_i=(u,c,t)$, $Q_{i2} \equiv d_i=(d,s,b)$ are
the up and down quarks, $l_{j1} \equiv \nu_{j}$
are the mass eigenstates of neutrinos and
$l_{j2} \equiv l_{j}=(e^{-}, \mu^{-}, \tau^{-})$
are the charged leptons.
The unitary matrices $A^{L,R}_{Q_a}$ and $A^{L,R}_{l_a}$ describe
the fermion mixing and diagonalize the mass matrices of quarks and
leptons.

The Higgs mechanism of the quark-lepton mass splitting needs, in
general, two scalar multiplets $\Phi^{(2)}$ and $\Phi^{(3)}$ (with
VEV $\eta_2$ and $\eta_3$) transforming according to the
representations   (1.2.1) and (15.2.1) of the group (\ref{eq:a1}).
The multiplet (15.2.1)
contains fifteen $SU_L(2)$ doublets:
\begin{eqnarray}
(15.2.1): \quad \,\,\,\,\,\,\,
\left ( \begin{array}{c} S_{1\alpha}^{(+)}  \\
         S_{2\alpha}^{(+)}\end{array} \right );
\left ( \begin{array}{c} S_{1\alpha}^{(-)}  \\
         S_{2\alpha}^{(-)}\end{array} \right );
\left ( \begin{array}{c} F_{1k}  \\
         F_{2k}\end{array} \right );
\left ( \begin{array}{c} \Phi_{1,15}^{(3)}  \\
         \Phi_{1,15}^{(3)}\end{array} \right ),
\label{eq:a2}
\end{eqnarray}
where $S^{(\pm)}_{a\alpha}$ and $F_{ak}$ (k=1,2...8) are the scalar
leptoquark and scalar gluons doublets. The scalar doublet
$\Phi^{(3)}_{15}$ is mixed with the (1,2,1) doublet $\Phi^{(2)}$
and gives the SM Higgs doublet $\Phi^{(SM)}$ (with SM VEV
$\eta=\sqrt{\eta_2^2+\eta_3^2}$) and an additional $\Phi'$
doublet. The scalar doublets~(\ref{eq:a2}) have the electric
charges
\begin{eqnarray}
Q_{em}:\quad \,\,\,\,\,\,\,
\left ( \begin{array}{c} 5/3  \\
         2/3\end{array} \right );
\left ( \begin{array}{c} 1/3  \\
        - 2/3\end{array} \right );
\left ( \begin{array}{c} 1  \\
         0 \end{array} \right );
\left ( \begin{array}{c} 1  \\
         0 \end{array} \right )
\nonumber
\end{eqnarray}
respectively.

In general case
the scalar leptoquarks $S_{2 \alpha}^{(+)}$ and
$\starup{S_{2\alpha}^{(-)}}$
with electric charge 2/3 are mixed and can be
written as superpositions
\begin{eqnarray}
S_{2 \alpha}^{(+)}&=&\sum_{m=0}^3 c_m^{(+)}S_m, \; \; \; \; \; \;
 \starup{S_2^{(-)}}=\sum_{m=0}^3 c_m^{(-)}S_m \label{eq:mixS}
\end{eqnarray}
of three physical scalar leptoquarks $S_1$, $S_2$, $S_3$ with
electric charge 2/3 and a small admixture of the Goldstone mode
$S_0$. Here $c^{(\pm)}_{m}$, $m=0,1,2,3$ are the elements of the
unitary scalar leptoquark mixing matrix, $|c^{(\pm)}_{0}|^2=\frac
{1} {3}g_4^2 \eta_{3}^{2}/m_V^2 \ll 1$, $g_4$ is the $SU_V(4)$
gauge coupling constant, $\eta_3$ is the VEV of the
(15,2,1)-multiplet and $m_V$ is the vector leptoquark mass.

The interactions of the scalar leptoquarks doublets 
$S^{(\pm)}_{a\alpha}$ with quarks and leptons can be written 
in the model independent form as 
\begin{eqnarray}
L_{S^{(+)}_1 u_i l_j} &=& \bar u_{i\alpha }  \Big [ ( h^L_+)_{ij}P_L + 
(h^R_+)_{ij}P_R  \Big ] l_{j} S_{1\alpha}^{(+)} + {\rm h.c.},
\nonumber\\
L_{S^{(-)}_1 \nu_i d_j} &=& \bar \nu_{i} \Big [ ( h^L_-)_{ij}P_L + 
(h^R_-)_{ij}P_R  \Big ] d_{j\alpha } S_{1\alpha}^{(-)} + {\rm h.c.},
\label{eq:Lsql} \\
L_{ S_m u_i \nu_j} &=& \bar u_{i\alpha} \Big [ 
(h^L_{1m})_{ij}P_L+(h^R_{1m})_{ij}P_R
\Big ] \nu_{j} S_{m\alpha} +{\rm h.c.} 
\nonumber\\
L_{ S_m d_i l_j} &=& \bar d_{i\alpha} \Big [ (h^L_{2m})_{ij}P_L + 
(h^R_{2m})_{ij}P_R \Big ] l_{j} S_{m\alpha} +{\rm h.c.} 
\nonumber
\end{eqnarray}
where $P_{L,R}=(1\pm\gamma_5)/2$ are the left and right projection
operators, $(h^{L,R}_\pm)_{ij}$ and $(h^{L,R}_{am})_{ij}$ are the
coupling constants, $i,j$ are the generation indexes.
Generally the interactions (\ref{eq:Lsql}) induce the scalar leptoquark decays 
\begin{eqnarray}
S_1^{(+)}\to u_il_j^+,\,\,\,S_1^{(-)}\to \nu_i\tilde d_j,\,\,\,
S_m\to u_i\tilde \nu_j,\,\,\,
S_m\to d_il_j^+.  
\label{eq:sdecay1} 
\end{eqnarray}

As a result of the Higgs mechanism of the quark lepton mass splitting 
the general form of Yukawa interaction in the MQLS model \cite{AD1,AD2,PovSm1}  
gives for the coupling  constants of (\ref{eq:Lsql}) the expressions
\begin{eqnarray}
(h^{L}_+)_{ij} &=&\sqrt{ 3/2} \frac{1}{\eta \sin\beta}
\Big [  m_{u_i} (K_1^LC_l)_{ij} -  (K^R_1)_{ik}m_{\nu_i} (C_l)_{kj} \Big ],
\nonumber \\
(h^{R}_+)_{ij} &=&-\sqrt{ 3/2} \frac{1}{\eta \sin\beta}
\Big [ (C_Q)_{ik}m_{d_k} (K_2^R)_{kj} - m_{l_j} (C_QK_2^L)_{ij}  \Big ],
\nonumber \\
(h^{L}_-)_{ij} &=&\sqrt{ 3/2} \frac{1}{\eta \sin\beta}
\Big [ (\krup{K^R_1})_{ik} m_{u_k} (C_Q)_{kj} - m_{\nu_j}
(\krup{K_1^L}C_Q)_{ij}  \Big ],
\label{eq:hsql} \\
(h^{R}_-)_{ij} &=&-\sqrt{ 3/2} \frac{1}{\eta \sin\beta}
\Big [( C_l \krup{K^L_2} )_{ij} m_{d_j}  -(C_l)_{ik} m_{l_k}
(\krup{K_2^R})_{kj}  \Big ], \nonumber\\
(h^{L,R}_{1m} )_{ij} &=& -\sqrt{ 3/2} \frac{1}{\eta \sin\beta}
\Big [  m_{u_i} (K_1^{L,R})_{ij} -
(K^{R,L}_1)_{ij}m_{\nu_j} \Big ] c_m^{(\pm)},
\nonumber\\
(h^{L,R}_{2m})_{ij} &=&-\sqrt{ 3/2} \frac{1}{\eta \sin\beta}
\Big [  m_{d_i} (K_2^{L,R})_{ij} -
(K^{R,L}_2)_{ij}m_{l_j} \Big ] c_m^{(\mp)},
\nonumber
\end{eqnarray}
where $m_{u_i},m_{d_i},m_{l_i},m_{\nu_i} $ are the masses of quarks,
of charge leptons and of neutrinos,
$\beta$ is $\Phi_a^{(2)}-\Phi_{15}^{(3)}$ mixing angle in MQLS model,
$tg\beta= \eta_3/\eta_2$,  
$C_Q = (A^L_{Q_1})^+ A^L_{Q_2}$ is the CKM-matrix,
$C_l = (A^L_{l_1})^+  A^L_{l_2}$ is the analogous matrix in
the lepton sector and 
$K^{L,R}_a = (A^{L,R}_{Q_a})^+ A^{L,R}_{l_a}$ are the mixing
matrices specific for the model with the four color quark-lepton
symmetry.

It is easy to see that among the coupling constants (\ref{eq:hsql}) 
there are the coupling constants 
\begin{eqnarray}
(h^{L}_{+})_{3j} &=&\sqrt{\frac{ 3}{2}} \frac{m_t}{\eta \sin\beta}
(K_1^{L}C_l)_{3j},\nonumber\\
(h^{L}_{-})_{i3} &=&\sqrt{\frac{ 3}{2}} \frac{m_t}{\eta \sin\beta}
(\krup{K_1^R})_{i3}(C_Q)_{33},\nonumber\\
(h^{L,R}_{1m})_{3j} &=&-\sqrt{\frac{ 3}{2}} \frac{m_t}{\eta \sin\beta}
(K_1^{L,R})_{3j}c_m^{(\pm)}\nonumber
\end{eqnarray}
which are proportional to the heavyest mass of $t$-quark 
and hence are the largest ones.

As a result among the decays (\ref{eq:sdecay1}) the scalar leptoquark decays 
\begin{eqnarray}
S_1^{(+)}\to tl_j^+,\,\,\,S_1^{(-)}\to \nu_i\tilde b,\,\,\,
S_m\to t\tilde \nu_j
\label{eq:sdecay2} 
\end{eqnarray}
into the quarks of the third generation and leptons 
occur to be the dominant ones.   


The calculation  for the case of 
$m_{l_j},m_{\nu_i}<<m_t$ and  $m_{\nu_i}<<m_b$
gives the following partial widths of the dominant 
modes (\ref{eq:sdecay2})
\begin{eqnarray}
\Gamma(S_1^{(+)}\to tl_j^+)=
\bar \Gamma_{S_1^{(+)}}(m_{S_1^{(+)}},m_t)
\frac{|(K_1^{L}C_l)_{3j}|^2}{sin^2\beta},
\label{eq:a4}\end{eqnarray}
\begin{eqnarray}
\Gamma(S_1^{(-)}\to \nu_i \tilde b)=
\bar \Gamma_{S_1^{(-)}}(m_{S_1^{(-)}},m_b)
\frac{|(\krup{K_1^{R}})_{i3}|^2|(C_Q)_{33}|^2}{sin^2\beta},
\label{eq:a5}\end{eqnarray}
\begin{eqnarray}
\Gamma(S_m\to t \tilde \nu_j)=
\bar \Gamma_{S_m}(m_{S_m},m_t)
\frac{|(K_1^{L})_{3j}|^2|c_m^{(+)}|^2 +
|(K_1^{R})_{3j}|^2|c_m^{(-)}|^2}{sin^2\beta},
\label{eq:a6}\end{eqnarray}
where   
\begin{eqnarray}
\bar{\Gamma}_{S}(m_S,m_Q) =
m_S\frac{3}{32\pi}(\frac{m_t}{\eta})^2 (1-\frac{m_Q^2}{m_S^2})^2  
\label{eq:gammas}
\end{eqnarray}
%
which we call
below as the reduced widths of the scalar leptoquarks  
 $S=S_1^{(+)}, S_1^{(-)}, S_m $.

Summarizing the partial widths (\ref{eq:a4})--(\ref{eq:a6}) 
over the generations and accounting for the unitaryty of the
matrices $K^{L,R}_1,C_l$ we obtain the next fermion mixing independent
expressions for the total decay widths
\begin{eqnarray}
\Gamma(S_1^{(+)}\to tl^+)&\equiv&\sum_j\Gamma(S_1^{(+)}\to tl_j^+) 
= \bar{\Gamma}_{S_1^{(+)}}(m_{S_1^{(+)}},m_t)\frac{1}{sin^2\beta} \, \, ,  
\label{eq:1e}\\
\Gamma(S_1^{(-)}\to \nu \tilde b)&\equiv&\sum_i\Gamma(S_1^{(-)}\to \nu_i \tilde b) 
= \bar{\Gamma}_{S_1^{(-)}}(m_{S_1^{(-)}},m_b)\frac{|(C_Q)_{33}|^2}{sin^2\beta} \, \, ,  
\label{eq:2e}\\
\Gamma(S_m\to t\tilde \nu)&\equiv&\sum_j\Gamma(S_m\to t\tilde \nu_j)
= \bar{\Gamma}_{S_m}(m_{S_m},m_t)\frac{k_m}{sin^2\beta} \, \, ,  
\label{eq:3e}
\end{eqnarray}
where                    
\begin{eqnarray}
 k_m = |c_m^{(+)}|^2 + |c_m^{(-)}|^2.  
\label{eq:km}
\end{eqnarray}

The interaction of the scalar gluons
with quarks can be written in the model independent form as 
\begin{eqnarray}
L_{F_1 u_i d_j} &=& \bar u_{i\alpha }  \Big [ ( h^L_{F_1})_{ij}P_L + 
(h^R_{F_1})_{ij}P_R  \Big ] (t_k)_{\alpha\beta}d_{j\beta} F_{1k} + {\rm h.c.},
\nonumber\\
L_{ F_2 u_i u_j} &=& \bar u_{i\alpha} \Big [ ( h^L_{1F_2})_{ij}P_L 
\Big ](t_k)_{\alpha\beta} u_{j\beta } F_{2k} + {\rm h.c.},
\label{eq:Lfqq}\\
L_{ F_2 d_i d_j} &=& \bar d_{i\alpha} \Big [ ( h^R_{2F_2})_{ij}P_R 
\Big ](t_k)_{\alpha\beta} d_{j\beta } F_{2k} + {\rm h.c.},
\nonumber
\end{eqnarray}
where $(h^{L,R}_{F1})_{ij}, (h^{L}_{1F1})_{ij}, (h^{R}_{2F1})_{ij} $ are 
the corresponding coupling constants and  $t_k, k=1,2\dots 8,$ are the generators 
of the $SU_c(3)$ group.
The interactions (\ref{eq:Lfqq}) induce the scalar gluon decays 
\begin{eqnarray}
F_1\to u_i\tilde b_j,\,\,\, F_2\to u_i\tilde u_j,\,\,\, F_2\to d_i\tilde d_j. 
\label{eq:fdecay1} 
\end{eqnarray}

From the general form of Yukawa interaction in the MQLS model \cite{AD1,AD2,PovSm1}  
the Higgs mechanism of the quark lepton mass splitting  
gives for the coupling constants of (\ref{eq:Lfqq}) the expressions
\begin{eqnarray}
(h^{L}_{F_1})_{ij} &=&\sqrt{ 3} \frac{1}{\eta \sin\beta}
\Big [  m_{u_i} (C_Q)_{ij} -  (K^R_1)_{ik}m_{\nu_k}(\krup{K_1^L}C_l)_{kj} \Big ],
\nonumber \\
(h^{R}_{F_1})_{ij} &=&-\sqrt{ 3} \frac{1}{\eta \sin\beta}
\Big [ (C_Q)_{ij}m_{d_i} -  (C_lK^L_2)_{ik}m_{l_k}(\krup{K_2^R})_{kj} \Big ],
\nonumber \\
(h^{L}_{1F_2})_{ij} &=&-\sqrt{ 3} \frac{1}{\eta \sin\beta}
\Big [  m_{u_i} (\delta)_{ij} -  (K^R_1)_{ik}m_{\nu_k}(\krup{K_1^L})_{kj} \Big ],
\nonumber \\
(h^{R}_{2F_2})_{ij} &=&-\sqrt{ 3} \frac{1}{\eta \sin\beta}
\Big [  m_{d_i} (\delta)_{ij} -  (K^L_1)_{ik}m_{l_k}(\krup{K_1^R})_{kj} \Big ] .
\label{eq:hfqq}
\end{eqnarray}

As seen among the coupling constants (\ref{eq:hfqq}) 
there are the coupling constants which are proportional to the $t$-quark mass  
\begin{eqnarray}
(h^{L}_{F_1})_{33} &=&\sqrt{ 3} \frac{m_t}{\eta \sin\beta}(C_Q)_{33},
\nonumber\\
(h^{L}_{1F_2})_{33} &=&-\sqrt{ 3} \frac{m_t}{\eta \sin\beta} 
\nonumber
\end{eqnarray}
and hence they are the largest ones.

As a result among the decays (\ref{eq:fdecay1}) the decays 
\begin{eqnarray}
F_1\to t\tilde b,\,\,\, F_2\to t\tilde t 
\label{eq:fdecay2} 
\end{eqnarray}
are the dominant ones.   

The calculation gives the following widths of the dominant modes (\ref{eq:fdecay2})
\begin{eqnarray}
\Gamma(F_1\to t\tilde b)=
m_{F_1}\frac{3}{32\pi}(\frac{m_t}{\eta})^2 
(1-\frac{m_t^2}{m_{F_12}^2})^2
\frac{|(C_Q)_{33}|}{sin^2\beta} 
\equiv \bar{\Gamma}_{F_1}(m_{F_1},m_t)\frac{|(C_Q)_{33}|}{sin^2\beta},  
\label{eq:4e}
\end{eqnarray}
\begin{eqnarray}
\Gamma(F_2\to t\tilde t)=
m_{F_2}\frac{3}{32\pi}(\frac{m_t}{\eta})^2 
(1-2\frac{m_t^2}{m_{F_2}^2})
\sqrt{1-4\frac{m_t^2}{m_{F_2}^2}}
\frac{1}{sin^2\beta}
\equiv \bar{\Gamma}_{F_2}(m_{F_2},m_t)\frac{1}{sin^2\beta},  
\label{eq:5e}
\end{eqnarray}
where the relation 
$m_b<<m_t$
has been taken into account.  

The decay widths (\ref{eq:1e})--(\ref{eq:3e}), (\ref{eq:4e}), (\ref{eq:5e}) 
depend on the masses of the decaying particles 
through the reduced widths $ \bar{\Gamma}_{S}(m_S,m_Q) , \bar{\Gamma}_{F}(m_F,m_Q) $ 
and on the mixing angle $\beta$  
and on the scalar leptoquark mixing parameters $k_m$. 
As mentioned above the indirect limits on the masses of the scalar leptoquarks 
under consideration are weak. 
The current data on the direct search for the leptoquarks set the
lower mass limits~\cite{PDG04}
\begin{eqnarray}
 m_{LQ}>242 \, \mbox{GeV}, \, 202 \, \mbox{GeV}, 148 \, \mbox{GeV} 
\label{dat:mass1}
\end{eqnarray}
 for the scalar leptoquarks of the first~\cite{GP98},
of the second~\cite{CDF97} and the of third~\cite{CDF00} generation with assuming 
the branching ratios $Br(lq) \equiv Br(LQ \to lq)$ of their quark - lepton decays 
to be $Br(e q)=1, Br(\mu q)=1, Br(\nu b)=1$ respectively. 
For the smaller values of $Br(lq)$ the  corresponding mass limits are weaker, 
for example  
\begin{eqnarray}
 m_{LQ}>204 (205) \, \mbox{GeV}, \, 160 \, \mbox{GeV} 
\label{dat:mass2}
\end{eqnarray}
for the scalar leptoquarks of the first~\cite{Abbott}~(\cite{Acosta}) and 
of the second~\cite{PDG04,Aball2} generation 
with $Br(e q)=0.5, Br(\mu q)=0.5$ respectively 
and 
\begin{eqnarray}
 m_{LQ}>79 (145) \, \mbox{GeV} 
\label{dat:mass3}
\end{eqnarray}
for the scalar leptoquarks of the first generation 
with $Br(e q)=0.0$~\cite{Abbott} ($Br(e q)=0.1$~\cite{Acosta}). 

As seen the scalar leptoquarks can have the masses of order of $200 \, GeV$ 
or slightly below in dependence on the the branching ratios of their fermionic decays. 
It is worth noting that because of their dominant decay modes (\ref{eq:sdecay2}) 
the scalar leptoquarks $S^{(+)}_1$, $S^{(-)}_1$, $S_m$ 
under consideration should be preferred as 
the third generation ones. Taking into account that the decay modes 
$ S_1^{(-)}\to \nu_i\tilde b $ of the scalar leptoquark $ S_1^{(-)} $ 
are the dominant ones with the total 
$ Br(S_1^{(-)}\to \nu \tilde b)  \equiv \sum_i Br(S_1^{(-)}\to \nu_i \tilde b) \approx 1 $ 
we have from (\ref{dat:mass1}) 
the lower mass limit $ m_{S^{(-)}_1}>148 \, \mbox{GeV} $ 
for the scalar leptoquark~$S^{(-)}_1$. 
As concerns the scalar gluons at the present time we have no direct experimental limits 
for their masses.           
Keeping in mind that the experimental bounds on the radiative corrections  
$S$, $T$, $U$ parameters of Peskin--Takeuchi allow the scalar leptoquark 
and scalar gluon doublets to lie below 1 TeV 
slightly prefferring their masses respectively of order of 400 GeV or below 
and of order of 800 GeV or below  
\cite{AD3,PovSm2} 
for the further numerical estimations 
we consider  the masses of these particle in sub TeV mass region.      
 
The Figure 1 shows the reduced widths $
\bar{\Gamma}_{S}(m_S,m_Q), \bar{\Gamma}_F(m_F,m_Q) $ 
 of the fermionic decays of the scalar leptoquarks 
$S = S^{(+)}_1, S^{(-)}_1, S_m$ and of the scalar gluons 
$F = F_1, F_2 $ as the functions of the masses $ m_S, m_F $ 
of the decaying particles in the mass region 200--1000 GeV. 
Here and below  we use the masses~\cite{PDG04} 
$m_t=174.3\pm5.1 \, GeV, \,\, m_b=4.25\pm0.25 \, GeV$ and 
$\eta=246 \, GeV.$ 
The curves a) and b) and c) correspond to the decays of the 
$ S^{(-)}_1 $ and $ S^{(+)}_1, S_m, F_1$ and $F_2$ respectively. 
The total widths of these decays for the more probable mass regions 
 are presented in Table 1.

As is seen from the the Fig.1 and from the Table 1 
the fermionic widths of the scalar leptoquark and gluon doublets can be  
of order of a few GeV with enhancing by the factor $1/sin^2\beta$. 
For example we obtain 
\begin{eqnarray}
\Gamma(S_1^{(+)}\to tl^+)&=&\Gamma(S_m\to t\tilde \nu)= 
0.2-5.8 \, (5.0-145.0) \, GeV,  
\label{eq:1en}\\
\Gamma(S_1^{(-)}\to \nu \tilde b)&=& 
2.2-7.5 \, (55.0-187.5) \, GeV,  
\label{eq:2en}\\
\Gamma(F_1\to t\tilde b)&=& 
5.8-14.1 \, (145.0-352.5) \, GeV,  
\label{eq:4en}\\
\Gamma(F_2\to t\tilde t)&=&
4.1-13.2 \, (102.5-330.0) \, GeV  
\label{eq:5en}
\end{eqnarray}
for $m_{S^{(+)}_1}, \, m_{S_m} = 200-500 \, GeV, \,
   m_{S^{(-)}_1} = 150-500 \, GeV, \, 
m_{F_1}, \, m_{F_2} = 500-1000 \, GeV $ 
and for $sin\beta=1 \, (0.2)$ 
with using the diagonal CKM matrix element $(C_Q)_{33}\approx 1$ and believing $k_m=1.$

\begin{table}[h]
\caption{ The total widths of the fermionic decays of the scalar leptoquarks
and of the scalar gluons
in dependence on the masses of the decaying particles.}
\bigskip
\begin{tabular}{|l|l|l|}\hline\rule[-2ex]{0ex}{5ex}
$\Gamma(S_1^{(+)}\to tl^+)$   & $(0.2-2.0-5.8)/\sin^2\beta $ \mbox{GeV}  &
$m_{S_1^{(+)}}=200-300-500 $ \mbox{GeV}
 \rule[-2ex]{0ex}{5ex}\\ \hline
$\Gamma(S_1^{(-)}\to \nu \tilde b)$   & $(2.2-4.5-7.5)|(C_Q)_{33}|^2/\sin^2\beta $ \mbox{GeV}  &
$m_{S_1^{(-)}}=150-300-500 $ \mbox{GeV}
 \rule[-2ex]{0ex}{5ex}\\ \hline
$\Gamma(S_m\to t \tilde \nu)$   & $(0.2-2.0-5.8)k_m/\sin^2\beta $ \mbox{GeV}  &
$m_{S_m}=200-300-500 $ \mbox{GeV}
 \rule[-2ex]{0ex}{5ex}\\ \hline
$\Gamma(F_1\to t\tilde b)$   & $(5.8-14.1)|(C_Q)_{33}|^2/\sin^2\beta $ \mbox{GeV}  &
$m_{F_1}=500-1000 $ \mbox{GeV}
 \rule[-2ex]{0ex}{5ex}\\ \hline
$\Gamma(F_2\to t\tilde t)$ & $(4.1-13.2)/\sin^2\beta $ \mbox{GeV}  &
$m_{F_2}=500-1000 $ \mbox{GeV}
 \rule[-2ex]{0ex}{5ex}\\ \hline
\end{tabular}
\end{table}

It should be noted that the decays (\ref{eq:sdecay1}), (\ref{eq:fdecay1}) 
with production of the quarks $Q$ of the first and of second generations 
are very suppresed by the factor $m_Q^2/m_t^2$ or by the squared matrix elements 
of CKM matrix $C_Q$. 
In the case of relatively small mass splittings $\Delta m$ inside the scalar doublets 
($\Delta m < m_W$) the weak decays 
\begin{eqnarray}
S\to S'W, \, F\to F'W  
\label{eq:sfwdecay}
\end{eqnarray}
of the heavyest components $S, \, F$ of the scalar doublets into the lightest ones $S', \, F'$ 
are forbidden and in this case 
 the decays (\ref{eq:sdecay2}), (\ref{eq:fdecay2})  are the doninant ones 
with the total branching ratios 
\begin{eqnarray}
Br(S_1^{(+)}\to tl^+) & \approx &  Br(S_1^{(-)}\to \nu \tilde b) \approx Br(S_m\to t \tilde \nu) \approx  1,  
\label{eq:brs2}\\
Br(F_1\to t\tilde b) & \approx & Br(F_2\to t\tilde t)\approx 1, 
\label{eq:brf2}
\end{eqnarray}
where 
$Br(S_1^{(+)}\to tl^+) \equiv \sum_j Br(S_1^{(+)}\to tl_j^+), \,$   
$Br(S_1^{(-)}\to \nu \tilde b) \equiv \sum_i Br(S_1^{(-)}\to \nu_i \tilde b), \, $  
$Br(S_m\to t \tilde \nu) \equiv \sum_j Br(S_m\to t\tilde \nu_j). $ 

In this case the simplest way for observation of the scalar leptoquarks 
$ S_1^{(+)}, \, S_1^{(-)}, \, S_m $ is the search for 
$te^+, \, t\mu^+, \, t\tau^+$-pairs which can be generated by the $ S_1^{(+)} $ decays  
with the total branching ratio $ Br(S_1^{(+)}\to tl^+) \approx 1 $ 
and the search for $\tilde b(t)$-quarks with energy missing  which can be generated by 
the $ S_1^{(-)} (S_m)$ decays 
with the total branching ratio $ Br(S_1^{(-)}\to \nu \tilde b) \approx 1 $ 
($ Br(S_m\to t \tilde \nu) \approx 1 $).   
The search for the decays (\ref{eq:sdecay2}) at LHC may be of interest 
and can result or in the observation of the scalar leptoquar doublets 
or in setting the new limits on their masses. 
As mentioned above the search for $\nu b$ pairs which has been performed at Tevatron 
set for the scalar leptoquark~$S^{(-)}_1$ 
the mass limit $ m_{S^{(-)}_1}>148 \, \mbox{GeV} $.    

In the case of $\Delta m > m_W$ the decays (\ref{eq:sfwdecay}) are also open. 
The analysys showes that the widths of the decays (\ref{eq:sfwdecay})  
can be comparable to the fermionic ones (\ref{eq:1en})-(\ref{eq:5en}) 
and in this case the relations (\ref{eq:brs2}), (\ref{eq:brf2}) are relevant 
only to the lightest components of the scalar doublets. 

As is seen from (\ref{dat:mass1})--(\ref{dat:mass3}) the scalar leptoquarks 
with the masses below the $t$-quark mass and with the sufficiently small 
$Br(e q)$ and $Br(\mu q)$ are not excluded. If the scalar leptoquarks 
 $ S=S_1^{(+)}, \, S_1^{(-)}, \, S_m $ are assumed to be lighter than $t$-quark  
($m_S < m_t$) then the decays  with production of $t$-quark in (\ref{eq:sdecay2}) 
are forbidden and in this case instead of the decays (\ref{eq:sdecay2}) 
the dominant decays are  
\begin{eqnarray}
S_1^{(+)}\to cl_j^+,\,\,\,S_1^{(-)}\to \nu_i\tilde b,\,\,\,
S_m\to b\l_j^+,\,\,\, S_m\to c\tilde \nu_j
\label{eq:sdecay3} 
\end{eqnarray}
with the total branching ratios 
\begin{eqnarray}
Br(S_1^{(+)}\to c l^+) & \approx &  Br(S_1^{(-)}\to \nu \tilde b) \approx 1, \,\,  
\label{eq:brs31}  \\
Br(S_m\to b l^+) & = & m_b^2/(m_b^2+m_c^2) \approx 0.9, \,\,                      
\label{eq:brs32}  \\
Br(S_m\to c \tilde \nu) & = & m_c^2/(m_b^2+m_c^2) \approx 0.1,   
\label{eq:brs33}  
\end{eqnarray}
where 
$Br(S_1^{(+)}\to c l^+) \equiv \sum_j Br(S_1^{(+)}\to c l_j^+), \,$   
$Br(S_1^{(-)}\to \nu \tilde b) \equiv \sum_i Br(S_1^{(-)}\to \nu_i \tilde b), \, $  
$Br(S_m\to b l^+) \equiv \sum_j Br(S_m\to b l_j^+), \, $
$Br(S_m\to c \tilde \nu) \equiv \sum_j Br(S_m\to c \tilde \nu_j). $ 
The search for the decays (\ref{eq:sdecay3}) at Tevatron with account of 
(\ref{eq:brs31})--(\ref{eq:brs33}) is of interest and could set the mass limits 
for the scalar leptoquarks $ S=S_1^{(+)}, \, S_m $ 
and the new mass limit for the scalar leptoquark~$S_1^{(-)}.$ 

In conclusion we resume the results of the work. The fermionic decays 
of the scalar leptoquark and scalar gluon doublets predicted by 
the four color symmetry with the Higgs mechanism 
of the quark-lepton mass splitting are investigated. 

The fermionic decays 
$ S_1^{(+)}\to tl^+_j, \,\,\,S_1^{(-)}\to \nu_i\tilde b,\,\,\,
S_m\to t\tilde \nu_j $
of the scalar leptoquarks 
$ S=S_1^{(+)}, \, S_1^{(-)}, \, S_m $
and those  
$ F_1\to t\tilde b,\,\,\, F_2\to t\tilde t $
of the scalar gluons $F=F_1, \, F_2$
are shown to be in the case of relatively small mass splittings $\Delta m$ 
inside the scalar doublets ($\Delta m < m_W$) 
 the dominant ones with the total branching ratios  
$ Br(S_1^{(+)}\to tl^+)  \approx   Br(S_1^{(-)}\to \nu \tilde b) 
\approx Br(S_m\to~t \tilde \nu)~\approx~1, \,\,  
Br(F_1\to t\tilde b)  \approx  Br(F_2\to t\tilde t)\approx~1. $ 
The widths of these decays are found to be of order of a few GeV 
for the masses of the decaying particles below 1 TeV. 

In the case of $m_S < m_t$ the scalar leptoquark decays 
$ S_1^{(+)}\to~cl_j^+ $, $S_1^{(-)}\to~\nu_i\tilde b $,  
$ S_m\to~b\l_j^+ $, $  S_m\to~ c\tilde \nu_j $ 
are shown to be the dominant ones with the total branching ratios    
$ Br(S_1^{(+)}\to~c~l^+) \approx $ $  Br(S_1^{(-)}\to~\nu~\tilde b)~\approx~1 $,    
$ Br(S_m\to~b~l^+) \approx 0.9 $,                        
$ Br(S_m\to~c~\tilde \nu)~\approx~0.1. $ 

The search for the considered decays at LHC and at Tevatron may be of interest.

\vspace{3mm} {\bf Acknowledgments}

The work was partially supported by the Russian Foundation for
Basic Research under grant 04-02-16517-a.

\newpage

{\Large\bf Figure captions}

\bigskip

\begin{quotation}

\noindent
Fig. 1. Reduced widths $\bar{\Gamma}_{\Phi}$ of the fermionic decays 
        of the scalar leptoquarks and of the scalar gluons as the functions 
        of the masses $m_{\Phi}$ of the decaying particles for  
           a)~$\Phi=S^{(-)}_1$, 
           b)~$\Phi=S^{(+)}_1, S_m, F_1 $, 
           c)~$\Phi=F_2$. 
\end{quotation}

\newpage
\begin{figure}[htb]
\vspace*{0.5cm}
 \centerline{
\epsfxsize=1.0\textwidth \epsffile{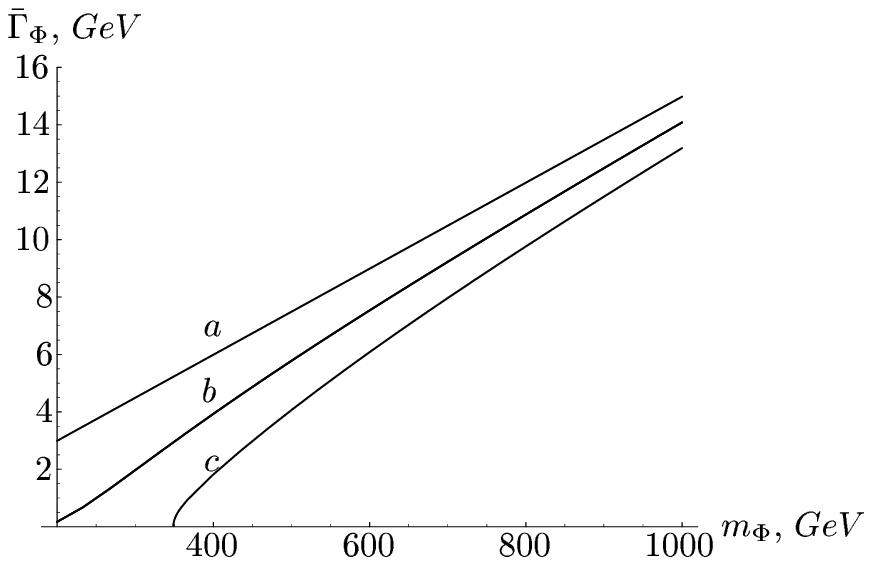}} \vspace*{1mm}
\label{fig:gammaSF}
\end{figure}
\vspace*{5cm}
\vfill \centerline{P.Yu.~Popov, A.V.~Povarov, A.D.~Smirnov, Modern Physics Letters A}
\centerline{Fig. 1}


\vspace{3mm}


\newpage
\vspace{-5mm}
\begin{thebibliography}{99}
\vspace{-2mm}
\bibitem{PS}
   J.~C.~Pati and A.~Salam,  Phys.~Rev.~D {\bf 10}, 275 (1974).
\bibitem{AD1}
   A.~D.~Smirnov, Phys.~Lett.~B {\bf 346}, 297 (1995).
\bibitem{AD2}
   A.~D.~Smirnov,  YaF. {\bf 58}, 2252 (1995).
   (Physics of Atomic Nuclei {\bf 58}, 2137 (1995).)
\bibitem{PovSm1}
  A.V. Povarov and A.~D.~Smirnov, YaF. {\bf 64}, 78 (2001).
  (Physics of Atomic Nuclei {\bf 64}, No.1, 74 (2001).)
\bibitem{AD3}
   A.~D.~Smirnov, Phys.~Lett.~B {\bf 513}, 237 (2002).
\bibitem{PovSm2} 
   A.V. Povarov and A.D. Smirnov
   YaF.  {\bf 66}, No.12, 2259 (2003).
   (Physics of Atomic Nuclei {\bf 66}, No.12, 2208 (2003).)
\bibitem{BBK1} 
  J. Bl\"umlein, E. Boos, A. Kryukov, Z.Phys. {\bf C76}, 137 (1997); hep-ph/9610408.
\bibitem{BBK2} 
  J. Bl\"umlein, E. Boos, A. Kryukov, Preprint DESY 97-067; hep-ph/9811271.
\bibitem{PDG04}
        Particle Data Group (S.~Eidelman et~al.), Phys.~Lett. {\bf B592}, 
         1 (2004).
\bibitem{GP98}
   C.Grosso-Pilcher et al. (The D0 and CDF Collaborations),
   hep-ex/9810015 .
\bibitem{CDF97}
   F.Abe et al.  (The CDF Collaboration),
   Phys.Rev.Lett. {\bf 78}, 2906  (1997).
\bibitem{CDF00}
   T.Affolder et al.  (The CDF Collaboration),
   Phys.Rev.Lett. {\bf 85}, 2056  (2000).
\bibitem{Abbott}
        B.~Abbott et al. (D0 Collaboration), Phys.~Rev.~Lett.
        {\bf80},~2051~(1998). 
\bibitem{Acosta}
        D.~Acosta et al. (CDF Collaboration), hep-ex/0506074.
\bibitem{Aball2}
        F.~Abe et al. (CDF Collaboration), Phys.~Rev.~Lett.
        {\bf 81},~4806~(1998).

\end{thebibliography}
\end{document}